\documentclass[alpha-refs]{wiley-article}
\usepackage{siunitx}
\usepackage{graphicx}
\usepackage[textwidth=8em,textsize=small]{todonotes}
\usepackage{amsmath}
\usepackage{natbib}
\usepackage{booktabs}
\usepackage{tabularx}
\usepackage{listings}
\usepackage{courier}

\papertype{Original Article}
\paperfield{Journal Section}

\title{Efficient design of geographically-defined clusters with spatial autocorrelation}

\abbrevs{ICC, intraclass correlation coefficient.}

\author[1\authfn{1}]{Samuel I Watson}
\affil[1]{University of Birmingham,
Birmingham,
United Kingdom.}
\corraddress{University of Birmingham,
Birmingham,
United Kingdom.}
\corremail{s.watson.1@warwick.ac.uk; s.i.watson@bham.ac.uk}

\fundinginfo{NIHR Global Health Research Unit on Improving Health in Slums.}

\runningauthor{S.I. Watson.}

\begin{document}

\maketitle

\begin{abstract}
Clusters form the basis of a number of research study designs including survey and experimental studies. Cluster-based designs can be less costly but also less efficient than individual-based designs due to correlation between individuals within the same cluster. Their design typically relies on \textit{ad hoc} choices of correlation parameters, and is insensitive to variations in cluster design. This article examines how to efficiently design clusters where they are geographically defined by demarcating areas incorporating individuals and households or other units. Using geostatistical models for spatial autocorrelation we generate approximations to within cluster average covariance in order to estimate the effective sample size given particular cluster design parameters. We show how the number of enumerated locations, cluster area, proportion sampled, and sampling method affect the efficiency of the design and consider the optimization problem of choosing the most efficient design subject to budgetary constraints. We also consider how the parameters from these approximations can be interpreted simply in terms of `real-world' quantities and used in design analysis.

% Please include a maximum of seven keywords
\end{abstract}

\section{Introduction}
Clusters form the basis of a variety of survey and experimental study designs. In a survey setting, multi-stage cluster sampling can offer a cost-effective alternative to simple random sampling when a study population is large or complete enumeration costly \citep{Fowler2012}. For a two-stage design, individual sampling units are often grouped into clusters on a geographic basis. These clusters are sometimes referred to as primary sampling units (PSU) or enumeration areas and may be constituted by political divisions like towns, villages, or census tracts, or they may be geographically defined for convenience. Cluster randomised trials (cRCT) are an experimental study design in which interventions are applied to whole clusters, which, as with survey designs, can also include geographically-defined areas \citep{Murray1998,Eldridge2012}. cRCTs are useful for evaluating processes applied at a `higher level' than the individual or where interaction between individuals within a cluster is unavoidable. In both survey and experimental studies with geographical clusters, individuals (or households or other sampling unit) from each cluster are typically enumerated to form a sampling frame and then sampled for inclusion and data capture. There is a variety of terminology used for cluster-based designs, often reflecting data collection or enumeration and sampling methodology. In this article we consider a cluster to be a set of individual sampling units grouped together because of their location in a geographically defined area for the purpose of enumeration, sampling, and data collection.

The potential benefit of a cluster-based study can be offset by the potential for reduced efficiency and the need for larger sample sizes to achieve the same level of precision. Should there exist correlation between individuals within the same cluster then the total amount of information a sample provides is reduced \citep{Skinner1986,Hooper2016}. One means of assessing this loss of information is to calculate the \textit{effective sample size}, which is the sample size of uncorrelated observations that would afford the same precision as the correlated sample. In many applications with clustered data, it is assumed that the covariance between observations in the same cluster is constant, such as in a linear mixed model framework where cluster means are assumed to be realisations of underlying normally-distributed latent variable \citep{Skinner1986,Murray1998,Hooper2016}. In this case the effective sample size is a function of the intraclass correlation coefficient (ICC), which is the proportion of the total variance attributable to the cluster-level. Sample sizes can then be inflated appropriately to provide a desired level of precision. However, this requires reliable estimates of the ICC.

There are relatively few studies providing estimates of ICCs from a wide range of settings for different variables. More commonly estimates exist from cRCTs rather than two-stage sampling schemes. For example, \citet{Campbell2005} estimate a large number of ICCs using data from `implementation science' cluster trials and report ICCs between 0.000 and 0.415, although the vast majority were below 0.1 with a median value below 0.05. A similar study of ICCs in cluster trials relating to heart failure reported ICCs between 0.026 and 0.052 \citep{Kul2014}. However, these trials do not use geographically-defined clusters, rather organisations or institutions like hospitals and schools. For two-stage sampling designs, \citet{Janjua2006} report ICCs of between 0.01 and 0.05 and \citet{Gulliford1999} report the same range in a study of the Health Survey for England. The values 0.01 to 0.05 are representative of those used in practice for designing both cluster trials and sampling schemes \citep{Hemming2017}. 

\begin{figure}
    \centering
    \includegraphics[width=\textwidth]{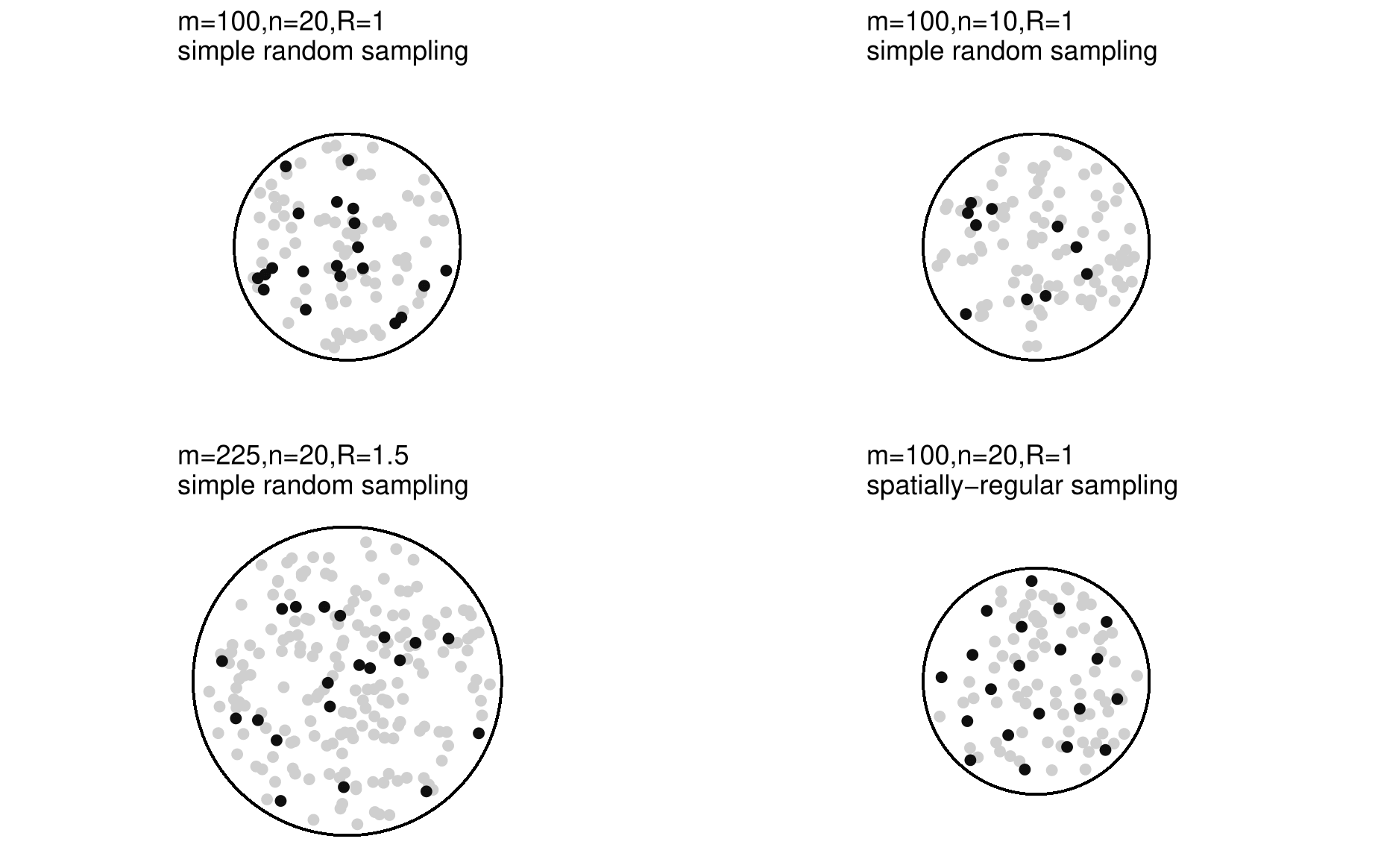}
    \caption{Illustrative example of variations in cluster design and sampling assuming a constant density of fixed locations, where $m$ is the number of locations enumerated (grey points), $n$ the number sampled (black points), and $R$ the radius of the cluster.}
    \label{fig:circles}
\end{figure}

The issue with using the ICC in this way for cluster-based designs is that it relies on the assumption of a constant covariance between individuals within a cluster and so it is unaffected by choices regarding the cluster design and sampling methodology. For geographically-defined clusters though, a likely explanation for within-cluster correlation is spatial autocorrelation, which exists for health (e.g.\citep{Tsai2009,MUNASINGHE1996,Lorant2001,Spielman2009}), economic (e.g.\cite{Basu1998,Molho1995}), and many other outcomes including research activity itself \citep{Elhorst2014}. The archetypal spatially autocorrelated outcome is infectious disease which requires the interaction of individuals for transmission \citep{Riley2007}. Given the nature of spatial autocorrelation, one can improve the efficiency of a cluster-based sampling scheme simply by enlarging clusters and sampling them more sparsely so that sampled units are further apart, for example (see Figure \ref{fig:circles}). However, this may require the generation of prohibitively large sampling frames, a reduction in cluster numbers, or `dilute' treatment effects from interventional studies. There is thus a trade-off between different cluster designs requiring statistical analysis. However, the reliance on an assumption of constant ICC for design purposes does not allow for this and little information exists on the sensitivity of an ICC to cluster design choices. In the absence of comprehensive estimates of ICCs for different sizes and types of cluster under different sampling schemes, it remains uncertain what effect altering the cluster design would have. Figure \ref{fig:circles} illustrates some different design choices.

If a model for spatial autocorrelation is assumed, then one can evaluate the effective sample size produced by a particular design on the basis of several empirically derived values including the distance between sampled points and information about the nature of the spatial correlation \citep{Griffith2005}. Enlarging clusters can increase the average distance between sampled points and improve efficiency, but if the sample size or number of individuals or households that can be enumerated to generate the sample frame is fixed then efficiency is lost by having fewer clusters. The cost to enumerate ever large clusters may also become prohibitive. Efficiency can be considered both in a purely statistical sense (effective sample size) or a cost-benefit sense (effective sample size per unit time or money). It is from this perspective that we examine cluster design in this article.

\section{The effective sample size under spatial autocorrelation}
For a sampling design for $N$ individuals, the loss of information due to any correlation between observations can be quantified by determining the effective sample size $N^* \leq N$, which is the sample size of uncorrelated observations that would afford the same precision as the correlated sample. This statistic is useful then to calculate the expected precision of a design in `power' or other types of calculation. 

Consider the linear model $Y = \mu + \Sigma^{\frac{1}{2}}e$ where $Y$ is an $n \times 1$ vector of observations, $\mu$ is the population mean, $e$ is an independent and identically distributed $N(0,\sigma^2_e)$ random variable, and $\Sigma$ contains information on the covariance structure so that the covariance matrix is given by $\Sigma\sigma^2_e$. In this context, \citet{Griffith2005} shows that the effective sample size is given by
\begin{equation}
\label{eq:effsamp}
    N^* = \frac{tr(\Sigma)}{\mathbf{1}'\Sigma\mathbf{1}}N
\end{equation}
where $\mathbf{1}$ is an $N \times 1$ vector of ones. For applications with simple clustering, such as a linear mixed model, the covariance matrix is typically assumed to have constant off-diagonal elements, $\sigma_{xx}^2$, for units in the same cluster and zero otherwise, and a variance of $\sigma^2$ on the diagonal. In this case Equation (\ref{eq:effsamp}) reduces to:
\begin{equation}
\label{eq:deseff}
    N^* = \frac{Jn}{1 + (n-1)\sigma_{xx}^2/\sigma^2}
\end{equation}
where $n$ is the average number of individuals in a cluster, $J$ is the number of clusters, and $\sigma_{xx}^2/\sigma^2$ is the intraclass correlation coefficient (ICC). The quantity $1 + (n-1)\sigma_{xx}^2/\sigma^2$ is sometimes also referred to as the `design effect' \citep{Skinner1986}. In the same way, one can derive the `design effect' for various other covariance matrices with compound symmetry such as designs with repeated measures and different specifications of temporal autocorrelation. \citet{Hemming2020} provide such a list in the context of cluster randomised trial design.

We can consider the effective sample size when the covariance between observations is instead determined by the spatial autocorrelation. The variogram is a widely used tool for estimating the relationship between distance and the degree of spatial autocorrelation \citep{Chiles2012}. In particular it is defined as 
\begin{equation}
    \gamma(d) = \frac{1}{2} E\left[ (Y(x+d) - Y(x))^2 \right]
\end{equation}
where $Y(x)$ is the values of the random variable $Y$ at location vector $x$ and $d$ is a spatial `lag' or distance. The empirical variogram can be defined by three parameters: the nugget, range, and sill. The sill ($C_1$) is the maximum value of $\gamma(d)$ and is equivalent to the total variance of $Y$, the range $r$ is the smallest value of $d$ at which $\gamma(d)$ reaches the sill, and the nugget ($C_0$) is the variance at scales shorter than the smallest observed distance or at the same location, for example the variance we might expect among individuals sampled in the same household or next-door neighbours (i.e. $\gamma(0)$). Then, as \citet{Griffith2005} describes, the variogram form of Equation (\ref{eq:effsamp}) is given by:
\begin{equation}
\label{eq:semio}
    N^* =
    \frac{Jn}{1+\frac{\rho}{n}\sum_{i=1}^N \sum_{j=1,j\neq i}^N f(d_{ij},r)}
\end{equation}
where $f(d_{ij}, r)$ is a given variogram model, $d_{ij}$ denotes the distance separating the locations of observations $i$ and $j$, and $\rho = (C_1-C_0)/C_1$.

When the parameter $\rho$ is zero then $C_0=C_1$ and there is no spatial autocorrelation, and when $\rho$ is one then $C_0$ is zero, which means there is no variance between observations present at the same location. Three different variogram functions are shown in Table \ref{tab:semifuncs}.

\begin{table}\small
\caption{Covariance functions and approximations to within-cluster mean covariance.}
    \centering
    \begin{tabularx}{\linewidth}{llll}
    \toprule
    Model & $f(d,r)$ & \multicolumn{2}{c}{Approx. mean covariance, $\Tilde{s}$} \\
    && Simple random   & Spatially-inhibited \\
    \midrule
        Gaussian & $ exp(-d_{ij}^2/r^2)$ & $1 - \frac{1}{(1+0.915 (r/R) ^ {2.071})}$ & $1 - \frac{1}{(1+0.876 (r/(\sqrt(R)+R/N) ^ {2.160})}$ \\
        Exponential & $ exp(-d_{ij}/r)$ &  $1 - \frac{1}{(1+0.764 (r/R) ^ {1.366})}$ &
        $-0.655 \text{tanh}(-0.795 (\frac{r}{\sqrt(R)+R/N})^{1.270})$\\
        \textit{K}-Bessel &  $ \frac{d_{ij}}{r} K_1(\frac{d_{ij}}{r})$ &  $1 - \frac{1}{(1+1.871 (r/R) ^ {1.603})}$ &
        $1 - \frac{1}{(1+1.829 (r/(\sqrt(R)+R/N) ^ {1.645})}$ \\
        \bottomrule
    \end{tabularx}
    
    \label{tab:semifuncs}
\end{table}

The effective sample size in this setup therefore depends only on the distance between sampled points and the parameters $r$ and $\rho$. It should be clear that as spatial autocorrelation increases $r \to \infty$, then $N^* \to N/(1-\rho (N-1))$, and in the presence of no spatial autocorrelation, $r \to 0$ then $N^* \to N$.

\subsection{Approximations to the ICC}
One can see that there is a correspondence between Equations (\ref{eq:deseff}) and (4) in that the equivalent `spatial' ICC is:
\begin{equation}
    ICC_{sp} = \frac{\rho}{n(n-1)}\sum_{i=1}^N \sum_{j=1,j\neq i}^N f(d_{ij},r)
\end{equation}
The term $s=\frac{1}{n(n-1)}\sum_{i=1}^N \sum_{j=1,j\neq i}^N f(d_{ij},r)$, and hence the locations of sampled units, is required for this calculation. However, prior to enumeration the locations of all units may be unknown and the precise pattern of sampling also uncertain. For practical applications an approximation to $s$, $\Tilde{s}$, using simpler values that are likely to be known \textit{ex ante} is required (\citet{Griffith2005} examine some similar approximations to the average covariance). The approximate effective sample size, for $J$ clusters of (mean) size $n$ would then be:
\begin{equation}
\label{eq:approxess}
    \Tilde{N}^* = \frac{Jn}{1+\rho \Tilde{s}(n-1)}
\end{equation}

Under simple random sampling, if we assume that the possible sampling locations are uniformly distributed across the area of interest, then the average interpoint distance remains the same for all sample sizes for a fixed area. The term $s$ therefore can be well approximated by some function of the radius of the area, $R$, assuming it is approximately circular. For cases where the area is better approximated by a quadrilateral then we use $\sqrt{A}$ where $A$ is the area of the area of interest. Our approximations to $s$ for simple random sampling are based on functions of the ratio $q_1=r/R$.

As an alternative to simple random sampling, one can use a spatially regular sampling method. Under such a scheme, points are selected at random, but in a way that ensures they are well spread out across the area. One particular design is the ``inhibited design with close pairs'', which can be used either for sampling from across a continuous area or from a set of fixed locations. A sample of size $n$ is randomly chosen so that no two locations are less that an inhibition distance $\delta$ apart. The sample is then supplemented with a set of ``close pairs'': $k$ sampled locations are randomly selected and then another location within distance $\eta$ is randomly sampled \citep{Chipeta2017}. This ensures that sampled locations are well dispersed across the area of interest. The close pairs ensure that information on shorter distances than $\delta$ is available (should it be required), although we assume no close pairs are included here. For a fixed area, a reduction in sample size will result in larger distances on average between sampled locations under a spatially-inhibited sampling design. The effect of this is to reduce the average covariance between sampled points. Approximations to $s$ for spatially regular sampling methods are therefore functions of both the area and sample size. In particular we consider the ratio:
\begin{equation*}
    q_2 = \frac{r}{\sqrt{R}+ R/n}
\end{equation*}

To generate an approximation we create a number of data sets of simulated data points. We first simulate 10,000 data sets of points uniformly distributed across a unit disc and vary $r$ between 0.001 and 2.000 and $n$ between 10 and 200. For each set of points we calculate $s$ for each variogram function $f$ from Table \ref{tab:semifuncs} and $q_1$. We then fit models of the form
\begin{equation*}
    s = g(q_1) + \epsilon
\end{equation*}
where $\epsilon \sim N(0,\sigma^2_\epsilon)$. For $g$ we consider a range of sigmoid functions including the logistic function, the hyperbolic tangent function, a cubic function, and the algebraic function $g(x) = 1 - 1/(1+\alpha q_1^\beta)$. The models were all estimated using maximum likelihood and their performance assessed by mean squared error. For the second data set the simulation parameters remain the same except points were simulated as regularly spaced across the unit disc. Models were specified in terms of $q_2$.

The third and fourth data sets were simulated in the same way as above, except that instead of a unit disc, points were either uniformly or regularly distributed across a quadrilateral with one side of length 1 and the other varying uniformly between 0.1 and 1. 

\begin{figure}
    \centering
    \includegraphics[width=\textwidth]{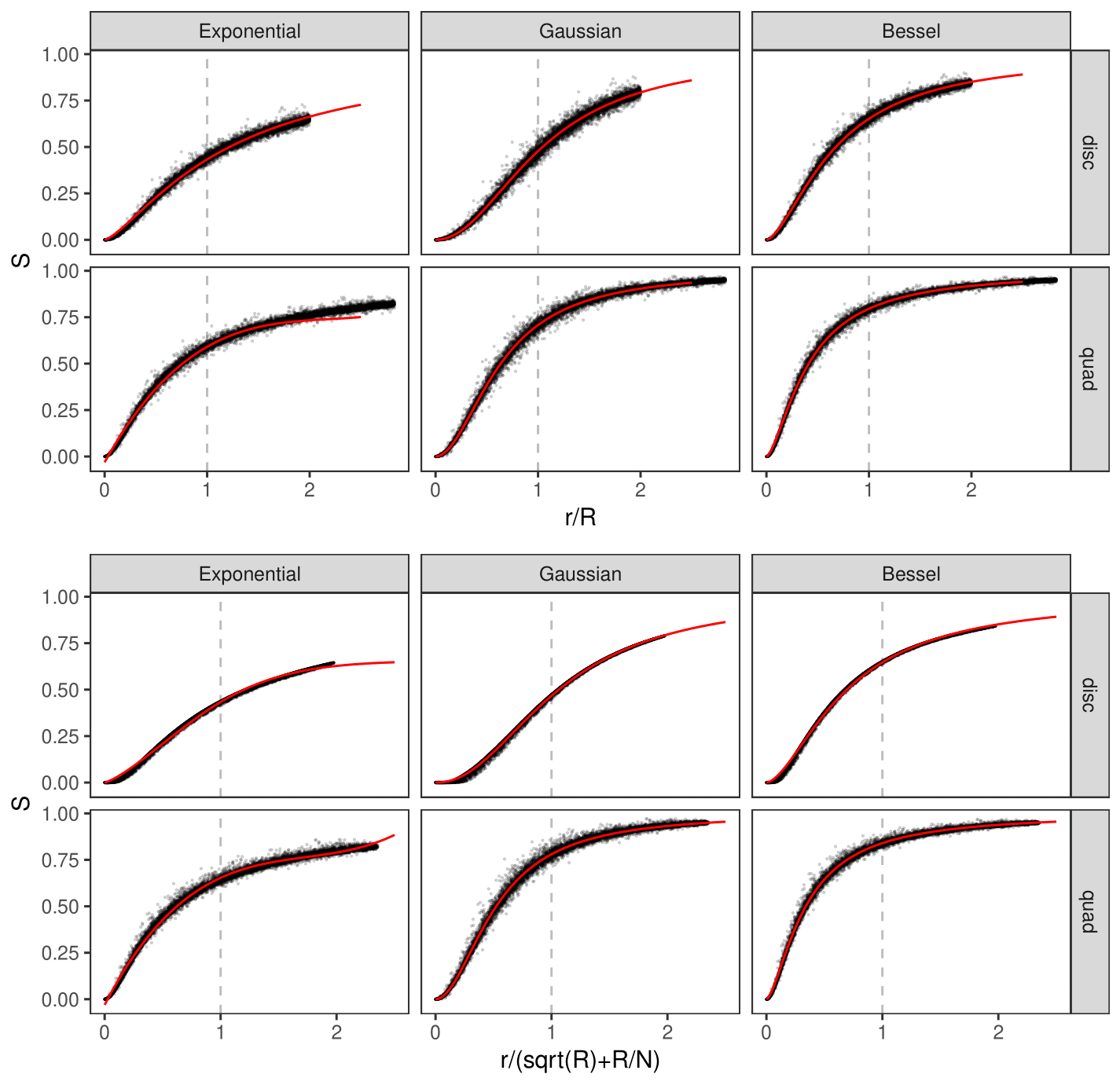}
    \caption{Simulated average covariance values $s$ (points) and best fitting approximations (red lines) for different semivariance functions (columns within panels), shape of the area (rows within panels), and uniformly (top panel) or regularly (bottom panel) distributed points.}
    \label{fig:approxicc}
\end{figure}

Figure \ref{fig:approxicc} shows the simulated data points and best fitting approximation and Table \ref{tab:semifuncs} reports the approximations for the disc (quadrilateral approximations are reported in Table XX,Appendix). Evident from these simulations is that the average covariance can be well approximated by the ratios specified above. Under simple random sampling, depending on the nature of the covariance function, there is little decline in the covariance with increasing $R$ until $R>r$, i.e. the radius or square root of the area is less than the range. When locations are well spread out, such as by using a spatially-regulated sampling method, then the average covariance is smaller. For example, for $r=0.5$ and $R=1$ and 20 sampled locations, the value for $\Tilde{s}$ with a Gaussian covariance function is 0.150, which reduces to 0.138 with 10 sampled locations. The reductions in average covariance are relatively small, but may translate into useful gains in effective sample size. If $\rho=0.3$ (so the equivalent ICCs are 0.0500 and 0.0414), then one cluster of size 20 where $s=0.150$ provides an approximate effective sample size of $\Tilde{N}^*=10.8$, whereas two clusters of size 10 and $s=0.138$ gives an approximate effective sample size of $\Tilde{N}^*=14.6$, a gain of approximately 20 percentage points in effective sample size per cluster.

\section{Simulation study of two-stage sampling schemes}
To illustrate the how choices over the size, number, and sampling scheme within clusters can affect the effective sample size we conduct a short simulation-based study. We describe a cluster-based sampling scheme using a number of parameters:
\begin{itemize}
    \item $J$: the number of clusters to be sampled;
    \item $m$: the average number of individuals/units to be enumerated within each cluster to form the sampling frame;
    \item $M=Jm$: the total number of enumerated locations;
    \item $p$: the proportion of the sampling frame to be sampled;
    \item $n = mp$: the average number of surveyed or enrolled individuals per cluster;
    \item $N = Jn = Jmp$: the total number of surveyed individuals.
\end{itemize}
Our interest lies in determining the effective sample size $N^*$. We assume that clusters are far enough apart so that there is no correlation between individuals in different clusters.

We examine two scenarios. First, $M$ and $p$ are fixed (and hence $n$) and only $J$ can be varied. The budget or time may only permit a certain number of locations and individuals to be enumerated and sampled, but the number of clusters can be flexibly varied. Since the number of locations that can be enumerated in a cluster depends on the number of clusters, as $J$ increases we assume that this implies a smaller cluster area. We simulate clusters of a disc shape and assume a density of 100 enumerated points per unit disc of area $\pi$. We vary the radius of the disc to give a specified number of points.

The second scenario is a scheme where $m$ is fixed but $J$ can be varied so that as $J$ increases, $M$ increases (as $m$ remains constant) and so the proportion of the enumerated locations in each cluster being sampled $p$ decreases. That is each cluster is the same size and the total number of individuals to be sampled is fixed so adding more clusters requires enumerating more locations and sampling fewer of them. 

For both scenarios we simulate points both uniformly and regularly distributed across the area to simulate random and spatially-regular sampling. Our target sample size is $N=200$. We conduct 10,000 simulations of each scenario and sampling scheme. For each iteration we generate $n$ points (either uniformly or regularly) according the parameters $m$ and $p$ and determine $n^*$ using Equation (\ref{eq:effsamp}) and each of the three semivariance functions, as well as the approximations described in the preceding section. We consider $\rho=0.01,0.1,0.5,0.9$ and $r = 0.1,0.5$.

\begin{figure}
    \centering
    \includegraphics[width=\textwidth]{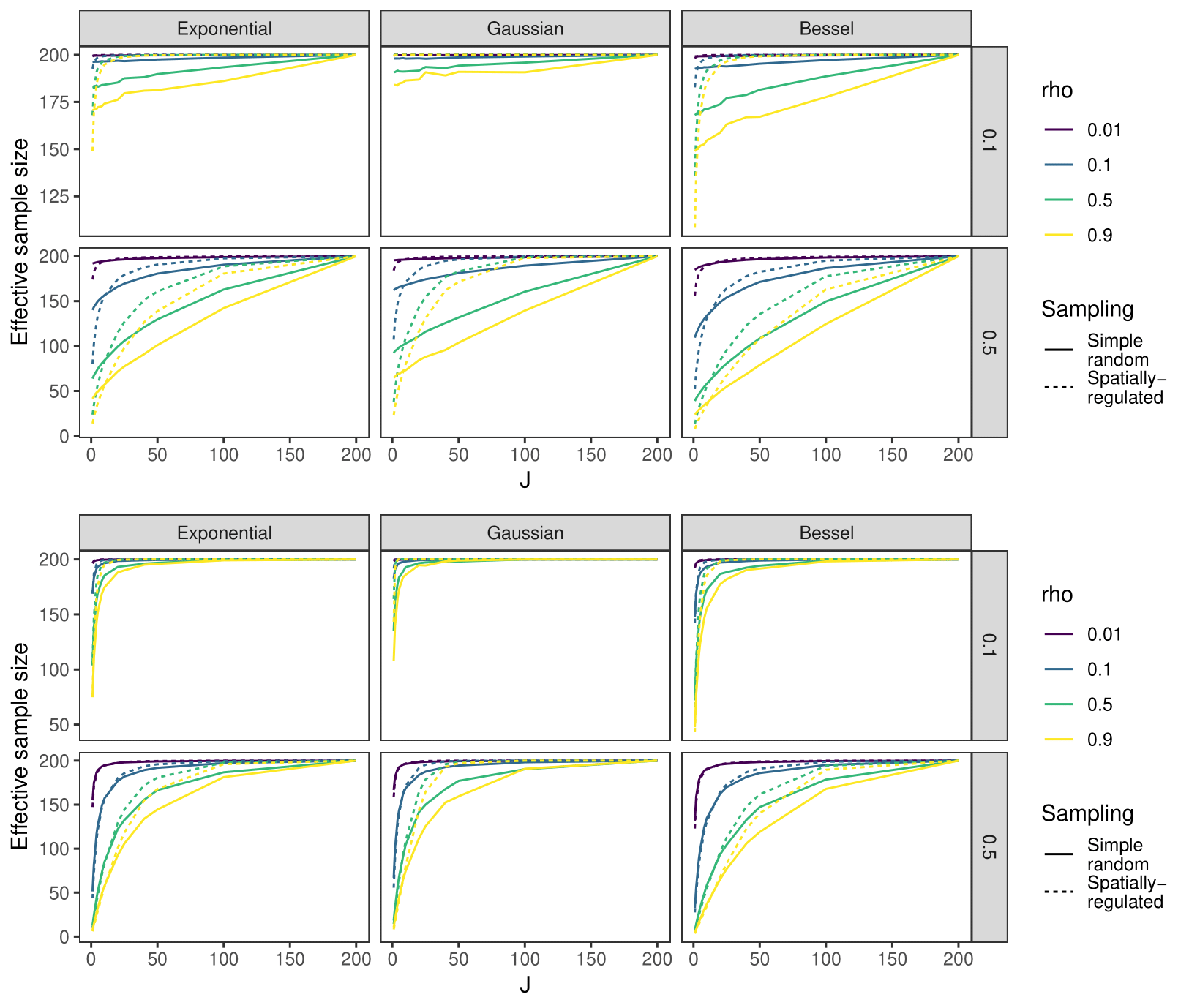}
    \caption{Results from the simulations showing effective sample size where $N=200$ for values of $\rho$ (color), sampling schemes (line types), values of $r$ (rows within panels), and semivariance functions (columns within panels). Top panel: fixed $M$ and $p$ with varying $J$. Bottom panel: fixed $m$ and varying $p$ and $J$.}
    \label{fig:sims}
\end{figure}

Figure \ref{fig:sims} reports the results of the simulations. These results can be interpreted in light of our approximations in the previous section. The top panel shows the effect of varying the number of clusters. Where the range is large ($r=0.5$), and hence the ratio $r/R$ is large and remains above one, there is an approximately linear relationship between the number of clusters and the overall effective sample size. However, where a spatially-regulated sampling scheme is used, increasing the number of clusters results in a rapid increase in the effective sample size as ratio $q_2$ declines below one. Where the range $r$ is low relative to the size of the cluster, there is little covariance between individuals and high effective sample sizes.

The lower panel of the figure reports the results of the simulations where the number of locations to be enumerated (and hence area) per cluster is fixed. An increase in the number of clusters results in fewer locations being sampled. There are exponential increases in the effective sample size under both sampling schemes as the number of clusters increases when $J$ is low indicating that the ratio $r/R$ is likely below one. The effective sample size is generally larger under spatially-regulated sampling, although there is little benefit in scenarios where $r$ or $\rho$ is small.

\section{Sampling parameters in practice and expert elicitation}
The efficiency of a geographic cluster-based study design depends (approximately) on three parameters: $\rho$, $r$, and $R$ (or $\sqrt{A}$). While the radius or area of the cluster area is determined by researchers, the values of the range and correlation parameters $r$ and $\rho$ are unlikely to be known and for most outcomes reliable estimates will not be available. However, both $\rho$ and $r$ (or transformations of them) can have relatively natural interpretations or be framed in terms of `real-world' variation, so that reasonable values could be specified by those with domain-specific knowledge. Expert elicitation methods could therefore provide useful bounds. This is an advantage over using the ICC, which is also unlikely to be known but does not have a useful `real-world' interpretation and there is little evidence of how it varies with cluster design. There are well specified methods for capturing expert beliefs about the distribution of variables or parameters \citep{OHagan1998}. In particular, we suggest the following interpretations for the cluster-design parameters:
\begin{itemize}
    \item $\sqrt{1-\rho}$: The ratio of the standard deviation between two very close neighbours to that of the whole population. One can follow procedures for the elicitation of a standard deviation. One might consider questions from which an estimate of standard deviation can be obtained, for example: `For a given value $a_1$ of a random variable, what interval $[a_2,a_3]$ would contain with probability $b$ the values of next door neighbours?' and `What interval $[a_4,a_5]$ would contain with probability $b$ the value of a randomly selected member of the population?' 
    \item $r$: The range parameter is the minimum distance over which two individuals have neglibible correlation. To elicit a value for this parameter, one can follow up the previous questions with: ``What is the minimum plausible distance over which the interval $[a_4,a_5]$ is plausible in the population?''
\end{itemize}
One can elicit either point estimates or probability densities. These estimates can be used in two ways. First, as a heuristic in designing the clusters. The clusters should aim to either have $R>r$ or if that is not possible to maximise the number of clusters. Second, the effective sample size can be determined. For example, if we define the approximate effective sample size to be (under an exponential variogram model and simple random sampling):
\begin{equation*}
    \Tilde{N}^*(r,\rho,R,J,n) = \frac{Jn}{1+\rho (n-1) \left( 1-(1+0.76 (r/R)^{1.37})^{-1} \right)}
\end{equation*}
then we can plug in point estimates or average over elicited probability densities:
 \begin{equation*}
     \Tilde{N}^*(R,J,n) = \int \int \Tilde{N}^*(r,\rho,R,J,n) h_r(r) h_\rho(\rho) dr d\rho
 \end{equation*}

\section{Dichotomous outcomes}
The analysis can be extended to studies that have dichotomous variables as outcomes. The major difference is in the interpretation and specification of $\rho$. If the outcome now takes values $Y\in \{0,1\}$ then we can rewrite the variogram in Equation (\ref{eq:semio}) as :
\begin{align}
 \gamma(d) &= \frac{1}{2}E\left[(Pr(Y(x+d)=1;Y(x)=0) + Pr(Y(x+d)=0;Y(x)=1)  )^2\right] \\
  &= Pr(Y(x+d)=1;Y(x)=0)
\end{align}
At the sill the two observations are independent so $\gamma(d)=Pr(Y(x)=1)(1-Pr(Y(x)=1))$, i.e. the variance of $Y$. At the nugget $\gamma(d)$ is equal to $Pr(Y(x+d)=0|Y(x)=1)Pr(Y(x)=1)$. Plugging these values in gives
\begin{equation}
    \rho = \frac{Pr(Y(x)=1|Y(x+d)=1) - Pr(Y(x)=1)}{Pr(Y(x)=1)(1 - Pr(Y(x)=1))}
\end{equation}
where the conditional probability is at the nugget (i.e. small $d$). Thus, to elicit a value for $\rho$ we need to ask:
\begin{enumerate}
    \item $Pr(Y(x)=1)$: What the expected probability of the outcome is across the whole population of interest;
    \item $Pr(Y(x)=1|Y(x+d)=1)$: If we observe a positive response, what is the probability that someone else living in the same household (or at the same location) would have a positive response.
\end{enumerate}
It is clear that if there exists no spatial autocorrelation then $\rho=0$, and if the autocorrelation is present then $Pr(Y(x)=1|Y(x+d)=1) > Pr(Y(x)=1)$ and $0<\rho\leq 1/Pr(Y(x)=1)$. 

\section{Optimising cluster-based study design}
Efficient cluster-based study design can also be seen as an optimisation problem of maximising effective sample size through the choice of $m$, $p$, and $J$ subject to financial or time constraints:
\begin{align}
    \text{Max. } \Tilde{N}^*(m,p,J) \text{ s.t. } & c_m M + c_n N \leq C \\
    & J \leq J_{max} \\
    & n \leq m 
\end{align}
where $c_m$ is the cost per enumerated location and $c_n$ is the cost per survey and where and $C$ is the total budget (where ``cost'' could be in time or money terms). The function to maximise depends on the choice of correlation function and sampling scheme, but, for example for the exponential correlation function under simple random sampling it is:
\begin{equation*}
    \Tilde{N}^*(m,p,J) = \frac{Jmp}{1+\rho (mp-1) \left( 1-(1+0.76 (r/(R))^{1.37})^{-1} \right)}
\end{equation*}
This optimisation problem can be solved relatively easily using statistical software. R Code to do this is provided in the Supplementary Information.

\section{Example}
Here we provide an example of the use of the design analysis presented in this article. We adapted a two-stage cluster sampling method for use in a household-based community survey on healthcare use in Kono District, Sierra Leone and Maryland County, Liberia. Census data on population numbers were almost a decade old and were thought to be an unreliable basis for a sampling frame particularly given recent events including the 2015 Ebola crisis. Moreover, population numbers were not captured at a level granular enough to enable stratification at village level or spatial sampling. Ground-truthing revealed that digital maps on the platform OpenStreetMap were out-of-date and between 10 and 20\% of structures were not accurately represented (either newly built or had been demolished). There did not exist funding to enumerate and map all the households in the districts, and so a two-stage scheme was adopted. An online application was generated that could be used to draw shapes over satellite images to create spatial data using the ArcGIS platform. Cluster borders could then be drawn around all structures and conurbations using the most recently available satellite images of the areas of interest. Areas were drawn with the aim of containing a pre-specified number of households and be bounded by features easily identifiable on the ground including roads, rivers, and tree lines. The number of households that could be sampled and interviewed was fixed, so the question remained as to how to determine the appropriate size of cluster to ensure reasonable efficiency while not requiring the enumeration of too large a number of households or travel to too many clusters.

\begin{figure}
    \centering
    \includegraphics[width=\textwidth]{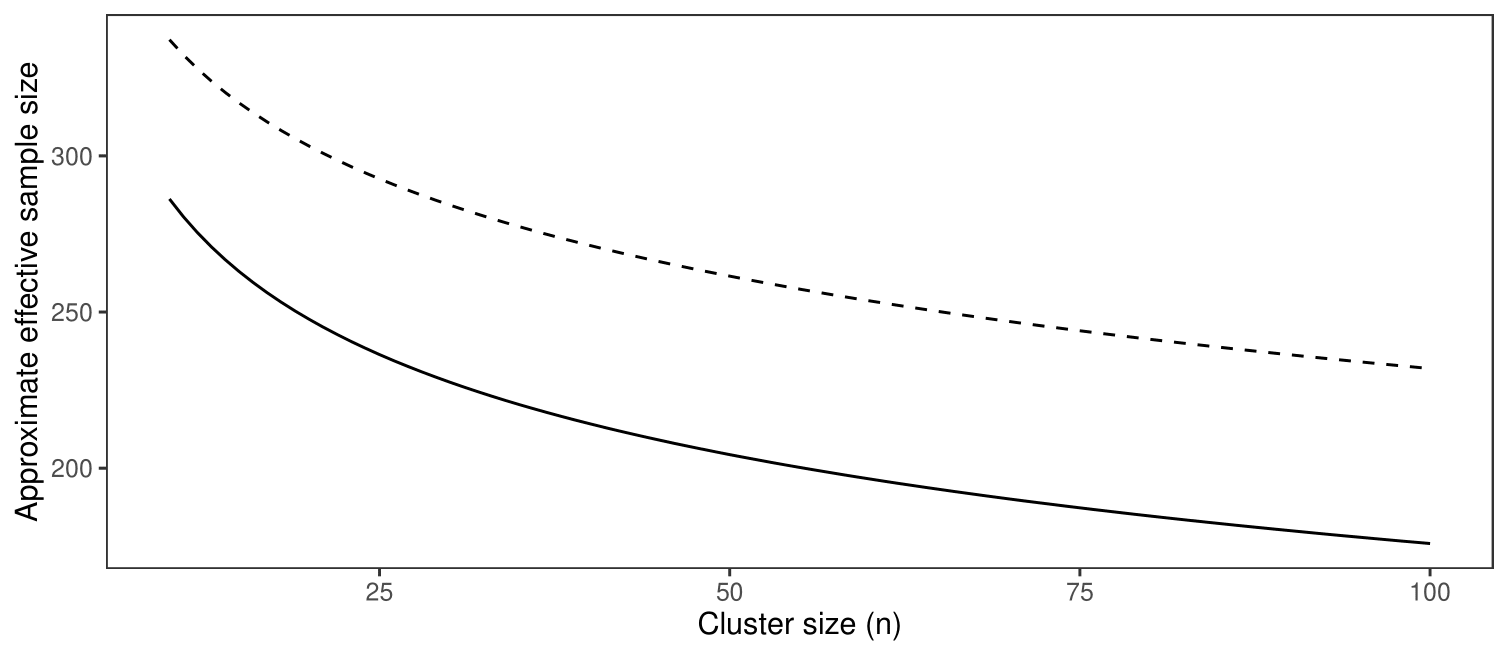}
    \caption{Approximate effective sample size for two strategies: survey all enumerated individuals (solid line) or enumerate double the number of individuals and sample half with simple random sampling (dashed)}
    \label{fig:sless}
\end{figure}

While there were multiple outcomes of interest, for the design analysis we focus here on estimating the proportion of the population who have seen a doctor, nurse, or clinical officer in an outpatient setting in the preceding 12 months. This variable is used as a filter for a number of following questions about the visits. Evidence from similar settings has suggested the outpatient consultation rate is around 0.5 visits per person-year, although the number of visits is not uniformly distributed in the population. $Pr(Y(x)=1)$ was therefore set at 0.4. For individuals living in the same household they might be more likely to attend if someone else visits a healthcare provider either because of possible shared sources of ill-health or for convenience (e.g. a parent taking a child). $Pr(Y(x)=1|Y(x+d)=1)$ was therefore set at 0.6. The range $r$ was set at 10m as these effects were not thought to extend beyond local neighbours relative to the population of the district outside of major pandemics such as Ebola or Covid-19. We assumed a constant density of households (approximately one per 3 square meters). Budget implications meant the total number of households that could be enumerated and surveyed was 500. The solid line in Figure \ref{fig:sless} shows the approximate effective sample size assuming an exponential correlation function where all enumerated households are surveyed ($p=1$). The most efficient strategy was to have the largest number of clusters possible which was around 30, giving 15 households per cluster and an approximate sample size of 260. As an alternative we examined doubling the cluster size and enumerating more households but taking a smaller sample to achieve the same approximate effective sample size (Figure \ref{fig:sless} dashed line) - this would reduce the number of households needing to be surveyed per cluster by around 20\% or 3 households in a cluster of size 15 but double the number being enumerated. However, this was less cost-effective. The optimum strategy identified by solving the optimisation problem in the previous section, assuming enumeration would take 10 minutes, surveying 50 minutes, and with a total `budget' of 20,000 minutes, provided $J=35$, $m=15$, and $p=0.72$, which gives an effective sample size of 274.

\section{Discussion}
The design of cluster-based studies is not an exact science. The evidence required to make precise \textit{ex ante} calculations of the precision afforded by a sample does not generally exist. Yet it is still important to ensure research designs are as efficient as possible. For studies based on geographically-defined clusters the standard method is to use the intraclass correlation coefficient (ICC) to estimate power, expected confidence interval width, or other measure of the information in a sample \citep{Skinner1986,Fowler2012,Hemming2017}. However, the values used for the ICC are often \textit{ad hoc}, based on commonly used values or occasionally evidence from studies of a similar design. While these values are representative of those estimated in empirical studies, it is not known how they should be varied with respect to clusters size, density, or method of sampling. Moreover, the ICC does not have a simply interpretable `real world' meaning, making expert elicitaiton type exercises difficult. In this article, we have explicated the efficiency of different geographical cluster design when the autocorrelation can be described in geostatistical terms. In particular, we show how effective sample size can be reasonably approximated with information on the range and parameter $\rho$. Small changes in cluster design can translate into relatively large changes in the effective sample size. Generally speaking the most efficient strategy is to maximise the number of clusters if the clusters are far enough apart to be independent. At the extreme there would be well-spaced out individuals in clusters of one. However, there is likely to be an upper limit on clusters for practical reasons including travelling and the number of locations that are far enough apart in the area of interest in which case methods to determine optimal designs are required. 

One can extend the analysis here to examine correlated clusters according to the distances between sampled units in each cluster. But \textit{inter}-cluster correlation would introduce further complications for interventional studies due to `contamination': intervening in one cluster has effects in another. In the presence of spatial autocorrelation this could be due to the underlying correlation or direct effects of the intervention over space. With appropriate geostatistical analysis these effects could be distinguished in experimental studies. The use of geostatistical methods in an experimental setting may provide useful avenues for research in this regard, however, analyses that take into space into account are exceedingly rare in cluster randomised trials. A systematic review of studies in 2017 only identified ten cluster trials that incorporated any kind of spatial analysis, and this was generally limited to allowing the intervention effect to vary linearly by distance from the intervention location \citep{Jarvis2017}. Model mis-specification could result in inefficient or biased estimators of treatment effects from these trials. Indeed, the interventional context provides an additional factor to consider for design in that larger clusters are more likely to have more `dilute' treatment effects if the intervention's effects decay by distance from a point source. This phenomenon also raises questions from cluster trials with geographically defined clusters as average treatment effects are clearly dependent on cluster design. There has been little development of the use of spatial statistical methods for experimental and interventional study designs, and further research is required.

For household surveys there has been little direct comparison of the most efficient sampling methods in settings where no sampling frame exists, particularly in terms of maximising the effective sample size for a given budget. \citet{Chao2012} compared two-stage methods with simpe random sampling from a complete sample frame and EPI sampling, a method based on surveying those close to randomly generated locations in the field, using a dataset of small businesses in South Africa. They suggested that a two-stage method was likely to be most cost-effective, however only considered a `compact segment sampling' approach with small clusters of a particular size in which all individuals were enumerated and surveyed. As this article demonstrates though, it is likely further improvements could be made by varying some of the design parameters. \citet{Milligan2004} compared EPI sampling and `compact segment sampling' in a field study and concluded that the latter method was preferred for its scientific validity but it was not strictly more efficient or cost-effective. The methods in this article permit a more precise approach to the decision over the design of such clusters and sampling strategies.

The analysis in this article reflects a more standard geostatistical sampling problem discussed widely elsewhere, that is how to efficiently sample locations across an area of interest in the presence of spatial autocorrelation \citep{Griffith2005,Manton2008}. The target for inference could be a population mean, the predicted prevalence of a disease and its spatial distribution, or the the parameters of a particular statistical model. A number of sampling schemes exist, which aim to evenly spread out a sample across an area. For example, a lattice-based design might overlay a hexagonal lattice over the area of interest and sample individuals at the vertices, which may be supplemented with simple random sampling in some of the cells to capture information on shorter-range correlation \citep{Yfantis1987}. We opted to consider a spatially-inhibited sampling design, which is more efficient than simple random sampling in the presence of spatial autocorrelation, but also is simple to implement for a set of fixed locations. Nevertheless, alternative sampling schemes may provide differign levels of efficiency for cluster-based designs. We also note that under spatially-regular sampling not all locations have an equal probability of being included in the sample as less densely packed areas will be more likely to be sampled, so appropriate probability weights should be calculated. 

There are of course a number of weaknesses to the approach proposed here. We rely on an assumption that sampling locations are uniformly distributed across the area of interest. This may not be a strong assumption for small, local neighbourhoods in an urban area in general, however, clusters that cross neighbourhood boundaries may have highly variable population density. The variogram functions used here may be relatively simple and not accurately describe the nature of spatial autocorrelation, however they are widely used for this purpose as they provide a good approximation to real-world variation. Indeed, the practical use of these methods is likely to be more complex than the scenarios presented here, for example, most clusters will not be regular discs or quadrilaterals. Nevertheless, these methods provide a more scientifically grounded approach to cluster design than currently used methods and as such greater validity.

The design of cluster-based studies commonly relies on ICCs, and sample size calculations are relatively straightforward under this approach. However, using ICCs in this way for geographically-defined clusters may lack scientific validity without reliable evidence on cluster-level variance. There also exists little evidence regarding how ICCs should be varied with respect to different design parameters like sampling method and cluster size. We have provided tools for more reliable cluster-design for surveys and experimental studies. One recommendation that arises from this study in particular is to use spatially regular sampling methods with geographically defined clusters as opposed to simple random sampling to improve efficiency.

\section{Citations and References}

\section*{conflict of interest}
None declared.

\printendnotes

% Submissions are not required to reflect the precise reference formatting of the journal (use of italics, bold etc.), however it is important that all key elements of each reference are included.
\bibliography{spaticc}

\appendix

\section{Approximations}

\begin{table}
    \centering
    \begin{tabular}{lccc}
    \toprule
         $g(x)$& Exponential & Gaussian & k-Bessel \\
         \midrule
         Unit disc, uniform distribution of points\\
         $\alpha \text{tanh}(\beta x^\gamma)$ & 0.0161 & 0.0005 & 0.0004\\ 
         $\alpha/(1+exp(-\beta x^\gamma)$ & 0.0144 & 0.0434 & 0.0544\\
         $1-1/(1+\alpha x^\beta)$ & \textbf{0.0003} & \textbf{0.0005} & \textbf{0.0003}\\
        $\alpha + \beta x+ \gamma x^2 + \delta x^3$ & 0.0003 & 0.0006 & 0.0004 \\
        \midrule
        Unit disc, regular spacing of points\\
        $\alpha \text{tanh}(\beta x^\gamma)$ & \textbf{0.0001} & 0.0001 & 0.0002\\ 
         $\alpha/(1+exp(-\beta x^\gamma)$ & 0.0117 & 0.0280 & 0.0589 \\
         $1-1/(1+\alpha x^\beta)$ & 0.0001 & \textbf{0.0001} & \textbf{0.0001}\\
        $\alpha + \beta x+ \gamma x^2 + \delta x^3$ & 0.0001 & 0.0002 & 0.0002\\
        \midrule
        Quadrilateral, uniform distribution of points\\
        $\alpha \text{tanh}(\beta x^\gamma)$ & 0.0004 & 0.0006 & 0.0005\\ 
         $\alpha/(1+exp(-\beta x^\gamma)$ & 0.0237 & 0.0276 & 0.0166\\
         $1-1/(1+\alpha x^\beta)$ & 0.0003 & \textbf{0.0003} & \textbf{0.0002}\\
        $\alpha + \beta x+ \gamma x^2 + \delta x^3$ & \textbf{0.0003} & 0.0006 & 0.0007 \\
        \midrule
        Quadrilateral, regular spacing of points\\
        $\alpha \text{tanh}(\beta x^\gamma)$ & 0.0004 & 0.0006 & 0.0005 \\ 
         $\alpha/(1+exp(-\beta x^\gamma)$ & 0.0237 & 0.0276 & 0.0166\\
         $1-1/(1+\alpha x^\beta)$ & 0.0003 & \textbf{0.0004} & \textbf{0.0002}\\
        $\alpha + \beta x+ \gamma x^2 + \delta x^3$ & \textbf{0.0003} & 0.0006 & 0.0007\\
         \bottomrule
    \end{tabular}
    \caption{Caption}
    \label{tab:approx1}
\end{table}

\begin{lstlisting}
# Find the optimum parameters for a sampling scheme. Inputs are:
# r = range
# R0 = average area occupied by one location
# rho = parameter rho
# cm = cost per enumerated location
# cn = cost per survey
# C = total budget
# Jlims = c(lower, upper) = the lower and upper limits to the number of clusters
# func = one of exp, gaus, or bess corresponding to exponential, Gaussian, or k-Bessel
# samp = one of simple or spatial for simple random sampling or spatially-regulated sampling
optimSamp <- function(r,R0,rho,cm,cn,C,Jlims,func,samp){
  if(!func%in%c('exp','gaus','bess')){
    stop("Function has to be one of 'exp', 'gaus', or 'bess' ")
  }
  if(!samp%in%c('simple','spatial')){
    stop("Sampling has to be either 'simple' or 'spatial' ")
  }
  
  a <- c(r,R0,rho)
  b <- c(cm,cn,C)
  
  if(func=="exp"){
    if(samp=="simple"){
      eval_f0 <- function(x,a,b){
        s <- -1*x[1]*x[2]*x[3]/(1+(x[2]*x[3]-1)*a[3]*
                                  (1 - 1/(1 + 0.76*(a[1]/(a[2]*sqrt(x[2])))^1.37)))
        return(s)
      }
    }
    if(samp=="spatial"){
      eval_f0 <- function(x,a,b){
        s <- -1*x[1]*x[2]*x[3]/(1+(x[2]*x[3]-1)*a[3]*
                                  (1 - 1/(1 + 0.88*(a[1]/
                                                      ( sqrt(a[2]*sqrt(x[2])) + 
                                                          a[2]*sqrt(x[2])/x[2]*x[3] ))^2.16)))
        return(s)
      }
    }
  }
  if(func=="gaus"){
    if(samp=="simple"){
      eval_f0 <- function(x,a,b){
        s <- -1*x[1]*x[2]*x[3]/(1+(x[2]*x[3]-1)*a[3]*
                                  (1 - 1/(1 + 0.92*(a[1]/(a[2]*sqrt(x[2])))^2.07)))
        return(s)
      }
    }
    if(samp=="spatial"){
      eval_f0 <- function(x,a,b){
        s <- -1*x[1]*x[2]*x[3]/(1+(x[2]*x[3]-1)*a[3]*
                                  -0.66*tanh(-0.80*(a[1]/
                                                      ( sqrt(a[2]*sqrt(x[2])) + 
                                                          a[2]*sqrt(x[2])/x[2]*x[3] ))^1.27 ))
        return(s)
      }
    }
  }
  if(func=="bess"){
    if(samp=="simple"){
      eval_f0 <- function(x,a,b){
        s <- -1*x[1]*x[2]*x[3]/(1+(x[2]*x[3]-1)*a[3]*
                                  (1 - 1/(1 + 1.87*(a[1]/(a[2]*sqrt(x[2])))^1.60)))
        return(s)
      }
    }
    if(samp=="spatial"){
      eval_f0 <- function(x,a,b){
        s <- -1*x[1]*x[2]*x[3]/(1+(x[2]*x[3]-1)*a[3]*
                                  (1 - 1/(1 + 1.83*(a[1]/
                                                      ( sqrt(a[2]*sqrt(x[2])) + 
                                                          a[2]*sqrt(x[2])/x[2]*x[3] ))^1.65)))
        return(s)
      }
    }
  }
  
  eval_g0 <- function(x,a,b){
    return(x[1]*x[2]*b[1] + x[1]*x[2]*x[3]*b[2] - b[3])
  }
  
  res1 <- nloptr(x0 = c(floor(Jlims[2]/2),20,1),
         eval_f = eval_f0,
         lb = c(Jlims[1],1,0),
         ub = c(Jlims[2],Inf,1),
         eval_g_ineq = eval_g0,
         opts = list("algorithm"="NLOPT_LN_COBYLA",
                     xtol_rel=1e-4),
         a=a,
         b=b)
  
  return(c(
    J = round(res1$solution[1],0),
    m = round(res1$solution[2],0),
    p = round(res1$solution[3],2),
    N = round(round(res1$solution[1],0)*round(res1$solution[2],0)*
                round(res1$solution[3],2),0),
    ess = -1*round(res1$objective,0)
  ))
}

# example call to function:
> optimSamp(10,15,0.5,30,50,25000,c(1,20),"exp","simple")
  J   m   p   N ess 
 20  16   1 320 215 
\end{lstlisting}

\end{document}